\documentclass[10pt]{article}
\usepackage{amsmath}
\usepackage{amssymb}
\usepackage[english]{babel}
\usepackage[latin1]{inputenc}
\usepackage{graphicx}
\usepackage{psfrag}
\usepackage{epsfig}
\newcommand{\nn}{\nonumber}
\numberwithin{equation}{section}
\title{\textbf{A covariant model of the electromagnetic current for the study of two-body scalar systems}}
\author{{\small Mario A. Acero$^1$, Maurizio De Sanctis$^{1,2}$, Carlos E. Sandoval$^1$}}
\date{\small{\textit{$^1$Universidad Nacional de Colombia, $^2$INFN, Sezione di Roma}}}
\begin{document}
  \maketitle%
\begin{abstract}
We present a procedure to derive a covariant electromagnetic current operator for a system made up by two scalars constituents. Using different wave functions we fitted their parameters to the experimental data of the pion form factor, obtainig great discrepancy at low momentum transfer. Introducing the Vector Meson Dominance corrective factor, we obtained a better fit to the data.
\end{abstract}
\section{Introduction}
In a previous work \cite{impulse} we introduced a covariant
impulse approximation method for the study of scattering processes
on composite systems. We briefly recall that in the impulse approximation one constituent interacts with the external field, while the others, the spectators, do not change their momenta in the interaction process. In the present paper we derive, by means of a gauge invariant procedure, the specific form of the
electromagnetic current operator for a system made up by two
scalars constituents.
\\
After discussing some general properties of the obtained current
operator, we study a numerical application to reproduce the
experimental behavior of the pion form factor, being aware that in
the present simplified ``scalar'' model all the spin effects are
neglected. Also, we verify that the approximations of the
procedure as such do not allow to reproduce the form factor at low
momentum transfer. To remedy to this failure we introduce a
suitable Vector Meson Dominance (VMD) corrective factor,
consistently with the other assumptions of the model. This VMD
factor takes into account, in a phenomenological way, all our
``ignorance'' about the structure of the bound state explored at
low momentum transfer.
\\
We point out that the main aim of the present work is to analyze
in a clear way our procedure for deriving a conserved current with
the simplest dynamical model, that is two scalar interacting
constituents.
\\
The same procedure, applied to the much more complex case
represented by the nucleon as a three-quark relativistic composite
system, is currently under examination and gives very good results
for the study of the electric and magnetic nucleon form factors \cite{MDeS}. Furthermore, in a subsequent work we shall study the pion form factor by considering a model with two spin $1/2$ -quark
antiquark- particles.
\\
The present work is organized as follows: In section 2, we analyze the dynamical
relativistic model and extract from the wave equation the
conserved current. Finally in section 3 we introduce the VMD factor and
draw some conclusions.
\section{The dynamical model and the gauge current}
\begin{figure}[htbp]
  \centerline{
    \psfrag{Q^2(GeV^2)}{$Q^2(GeV^2)$}
    \psfrag{F(pion)}{\small{$F_{\pi}$}}
    \psfrag{F(pi)_1}{\small{$F_{\pi_1}$}}
    \psfrag{F(pi)_2}{\small{$F_{\pi_2}$}}
    \psfig{figure=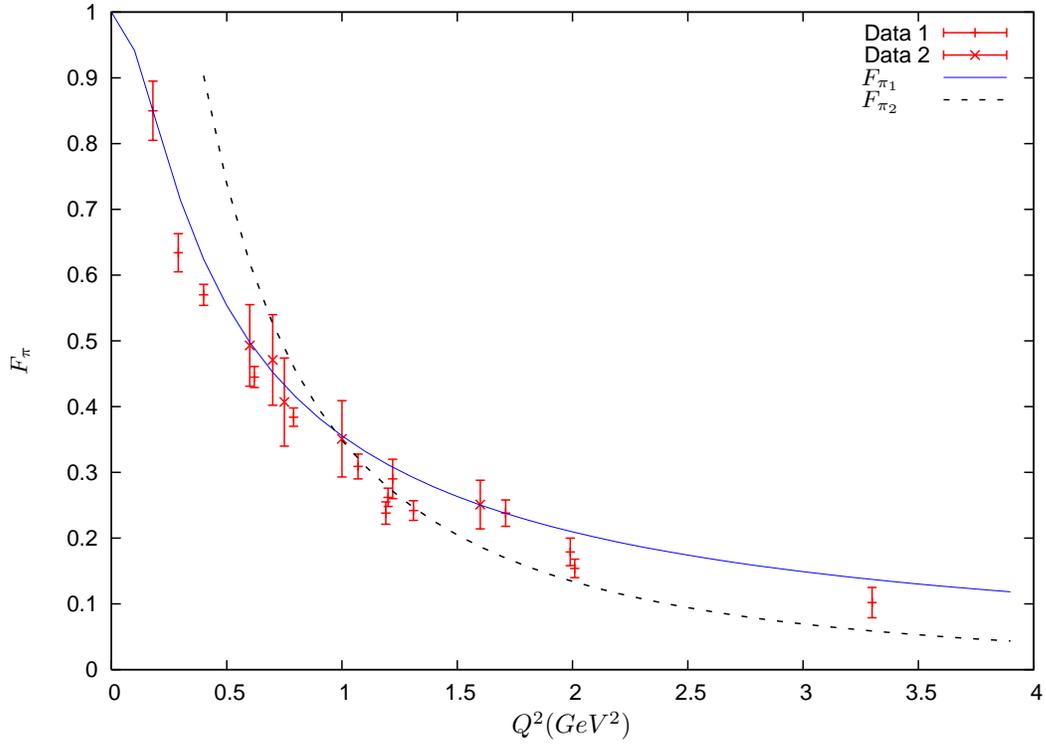,width=10cm,angle=-90}
  }
  \caption{\small{Pion form factor with ($F_{\pi_1}$) and without ($F_{\pi_2}$) VMD, compared with experimental data. For $F_{\pi_1}$, we have $a=0.70$, $b=0.55$, $C=0.76(GeV/c)^{-3/2}$, $\alpha=0.9(GeV/c)$ and $\beta=0.6(GeV/c)$, in equation (\ref{3.4}); in addition, for the VMD factor (equation (\ref{3.1})), we set $m_V=m_{\rho}=0.77(GeV/c^2)$, $\beta_V=0.865$ and $n=2$. For $F_{\pi_2}$, we use the standard Hulten wave function, with $\alpha=0.1(GeV/c)$ and $\beta=0.4(GeV/c)$. (Data 1: ref. \cite{bebek}, Data 2: ref. \cite{volmer})}}\label{fig6}
\end{figure}

We firstly introduce the relativistic wave-equation for the
two-body system. In its rest frame (RF) it has the standard form
of a positive energy relativistic integral equation \cite{QFT}
\begin{equation}\label{2.1}
[M-2E(p)]\varphi(\vec p) = \int d^3p' W(\vec
p,\vec{p'})\varphi(\vec{p'}),
\end{equation}
where
\begin{equation}\label{2.2}
\vec p = \vec{p_1^*} = -\vec{p^*_2}
\end{equation}
represents the relative momentum of the constituents; hereafter, the asterisk denotes the quantities referred to the RF;
\begin{equation}\label{2.3}
E = E_1^* = E_2^* = \sqrt{m^2+\vec p^2}
\end{equation}
represents the kinetic energy of the constituents, being $m$ their
mass; finally, $W(\vec p,\vec{p'})$ denotes the interaction
operator.
\\
In a generic reference frame that is chosen for the observation of
the scattering process, called Observation Reference Frame (ORF),
the 4-momenta, on-shell, of the two constituents $p_1$,
$p_2$, are obtained with a Lorentz transformation on the
corresponding quantities of the RF. Standard calculations give the
following relation:
\begin{equation}\label{2.4}
p_1+p_2 = \frac PM\bar E,
\end{equation}
where
\begin{equation}\label{2.5}
\bar E = 2E(p) = 2\frac{Pp_1}{M}.
\end{equation}
In the previous equations the total observable 4-momentum of the
system $P$ is introduced with $P^2 = P_{\mu}P^{\mu} = M^2$.
Equation (\ref{2.4}) represents the main assumption of the present
model that is based on the use of on-shell momenta as dynamical
spatial variables.
\\
Also, by means of the last two equations it is easily shown that
in an ORF, the on-shell 4-momentum $p_2$ can be considered as a
function of $p_1$ and $P$. In consequence we can make for the explicitly covariant 
wave-equation in an ORF the following hypothesis:
\begin{equation}\label{2.6}
[M-2\frac{Pp_1}{M}]F\varphi = \int \frac{d^3p'_1}{E(p'_1)}
V(p_1,p'_1,P)F'\varphi',
\end{equation}
being $F$ an \underline{invariant} function: $F=F(p_1,P)$.
Calculating the previous equation in the RF, the requirement of
hermiticity of the interaction leads to
\begin{equation}\label{2.7}
F(p_1,P) = \sqrt{E(p)} = \sqrt{E_1^*} = \left(\frac{Pp_1}{M}\right)^{1/2},
\end{equation}
and, comparing with equation (\ref{2.1}), one finds that the interaction
of the ORF equation is related to that of the RF by the following relation:
\begin{equation}\label{2.8}
\frac1F V(p_1,p'_1,P)\frac{1}{F'} = W(\vec{p_1^*},\vec{p_1^{*'}}) = W(\vec{p},\vec{p'}).
\end{equation}
Also
\begin{equation}\label{2.9}
\varphi(p_1,P) = \varphi(\vec{p_1^*}(p_1,P)).
\end{equation}
The last equation shows that the rest frame momentum must be expressed as a
function of the ORF momentum $p_1$ and of the total 4-momentum $P$
by means of a standard Lorentz transformation.
\\
Starting from the covariant wave equation in the ORF of equation
(\ref{2.6}), we derive the gauge invariant conserved current by
means of the following procedure.
\begin{enumerate}
\item We specialize equation (\ref{2.6}) to the initial state,
that is $P=P_I$.
\item To obtain the current related to the constituent 2, being
the constituent 1 the spectator, one should multiply by its
charge, $e_2$. However, in the example under examination, due to the symmetries of the wave
function, to derive the total current one can directly multiply by
the total charge of the system, denoted by $Q$.
\item We multiply by $F(p_1,P_F)\varphi^+(p_1,P_F)$ and perform
the covariant integration over $\int \frac{d^3p_1}{E(p_1)}$.
\item We repeat this procedure for the h.c. wave function of the final state.
\end{enumerate}
Subtracting the two results we obtain
\begin{align}\label{2.10}
Q&\int \frac{d^3p_1}{E(p_1)} \varphi^+(p_1,P_F)F(p_1,P_F)
\left[\Delta[M]-2\Delta\left[\frac{P^{\nu}}{M}\right]
p_{1\nu}\right]F(p_1,P_I)\varphi(p_1,P_I)\nn \\ \nn\\
&= Q\int \frac{d^3p_1}{E(p_1)}\frac{d^3p'_1}{E(p'_1)}
\varphi^+(P_F,p_1)F(P_F,p_1)\Delta[V(p_1,p'_1,P)]F(P_I,p_1)\varphi(P_I,p_1),
\end{align}
where the notation
\begin{equation}\label{2.11}
\Delta[O(P)]=O(P_F)-O(P_I)
\end{equation}
has been introduced. In particular, we point out that in all the
$\Delta$ quantities in equation (\ref{2.10}) the momentum transfer
$q^{\mu}=(P_F^{\mu}-P_I^{\mu})$ can be factorized. We have
\begin{equation}\label{2.12}
\Delta[M] = q_{\mu}\frac{2K^{\mu}}{M_F+M_I},
\end{equation}
\begin{equation}\label{2.13}
\Delta\left[\frac{P^{\nu}}{M}\right] = q_{\mu}
\left[\frac12\frac{M_F+M_I}{M_FM_I}g^{\mu\nu} -
\frac{2K^{\mu}K^{\nu}}{M_FM_I(M_F+M_I)}\right],
\end{equation}
with
\begin{equation}\label{2.14}
K= \frac12(P_F+P_I).
\end{equation}
Also, $\Delta[V(p_1,p'_1,P)]$ can be expressed in a similar way as
it will be shown in a subsequent work. We obtain a current $J_{FI}^{\mu}$ that satisfies the gauge condition
\begin{equation}\label{2.15}
q_{\mu}J^{\mu}_{FI} = 0.
\end{equation}
In this paper we study the terms obtained from the LHS of equation
(\ref{2.10}) that represent the ``kinetic'' contributions. We
have:
\begin{align}\label{2.16}
J_{FI}^{(k)\mu} &= Q\int \frac{d^3p_1}{E(p_1)}\varphi^+(P_F,p_1)F(P_F,p_1)\\
&\times \left[\frac{2K^{\mu}}{M_F+M_I} +
\left(\frac12\frac{M_F+M_I}{M_FM_I}g^{\mu\nu}-\frac{2K^{\mu}K^{\nu}}{M_FM_I(M_F+M_I)}\right)
p_{1\nu}\right]F(P_I,p_1)\varphi(P_I,p_1),\nn
\end{align}
note that the matrix-element of
the current of the last equation is of the same kind of that
introduced in our previous work -see equation (3.8) of
\cite{impulse}-. In other words, the contribution to the current
coming from the kinetic energy of the covariant wave equation
gives rise to a term of impulse approximation form, where the spectator's 4-momentum $p_1$ remains unchanged. The
operator of the interacting particle that is represented in the present work by the term in the parenthesis of equation
(\ref{2.16}), was generically denoted as $\hat O_2$ in equation (3.8) of \cite{impulse}. Furthermore, standard but tedious calculations show that, in the present model, this operator, in the limit of free
constituents, is proportional to $p_{2F}^{\mu}+p_{2I}^{\mu}$ as it
must be for scalar particles. Finally, the factors $E^{-1}(p_1)$, $F(P_F,p_1)$, $F(P_I,p_1)$ represent the covariant normalization factors of the wave function, also introduced in \cite{impulse}.
\\
In the following we study the case of elastic scattering, that is,
$M_F=M_I$; the elastic form factor is related to the
matrix-elements of the current by the following equation:
\begin{equation}\label{2.17}
J_{FI}^{\mu} = \frac{K^{\mu}}{M}F(q^2),
\end{equation}
that will be used to determine $F(q^2)$.
\\
For the calculations the Breit reference frame will be used. In
this frame, one has
\begin{align}
P_{F/I} &= \left(\left[M^2+\frac{q^2}{4}\right]^{1/2},\pm\frac{\vec q}{2}\right)\label{2.18} \\
q &= (0,\vec q)\label{2.19} \\
K &= \left(\left[M^2+\frac{q^2}{4}\right]^{1/2},\vec 0\right).\label{2.20}
\end{align}
Also, the momentum transfer $\vec q$ is taken along the $\hat z$ axis. In consequence, for an elastic scattering, in the Breit Frame, the matrix elements of the current take the form
\begin{equation}\label{2.21}
J_{FI}^{(k, Breit)\mu}= Q \int \frac{d^3p_1}{E(p_1)} \varphi^\dag(P_F,p_1)F(P_F,p_1) I^\mu F(P_I,p_1)\varphi(P_I,p_1),
\end{equation}
where, from equation (\ref{2.16}), $I^{\mu}$ has the form
\begin{equation}\label{2.22}
I^0=\frac EM \left[1+\frac 12 \frac{q^2}{EM^2} E(p_1)\right],
\end{equation}
\begin{equation}\label{2.23}
\vec I = -2\frac{\vec p_1}{M}.
\end{equation}
Note that, even though equation (\ref{2.16}) only represents the kinetic contribution to the current, this current is conserved, in elastic scattering on a scalar composite system. This property can be proved by means of invariance under parity transformations.
\\ 
To this aim let us consider the operator $p_1^z$, which is the only one appearing in the current conservation relations. It satisfies the following relation under parity transformation:
\begin{equation}\label{2.24}
P^\dag p_1^z P=-p_1^z.
\end{equation}
Furthermore, in this case, that is, elastic scattering on a scalar state in the Breit frame, the parity transformation interchanges the initial and final states:
\begin{equation}\label{2.25}
PF(P_I,p_1)\varphi(P_I,p_1)=F(P_F,p_1)\varphi(P_F,p_1).
\end{equation}
The last two equations are sufficient to prove current conservation, that, due to the covariance of the model, holds in all the reference frames.
\\\\ 
Considering the numerical results for the form factor, we found that with standard wave-functions as gaussian, Hulten, or the modified Hulten that will be introduced later, this simple scalar model reproduces the high momentum transfer behavior of the pion form factor, while the low momentum transfer region cannot be correctly reproduced. An example is shown in figure \ref{fig6}. 
This problem can be related to the highly non perturbative character of the system when explored at low momentum transfer, i.e. at large spatial distance.
\\ 
On the other hand, the model, being essentially a relativistic impulse approximation, explains correctly the ``partonic'' behavior at high momentum transfer, i.e. at short spatial distance.
\\ 
A phenomenological way to improve the model is to introduce the so called Vector Meson Dominance, as it will be explained in the next section.
\section{The Vector Meson Dominance corrective factor}
The Vector Meson Dominance (VMD) is an old model that has been extensively used for the study of the electro-weak interactions of hadronic systems. As for the electron scattering processes, the initial photon interacts not only in a direct way with the hadronic system, but also by means of an intermediate meson ($\rho, \omega, \phi, ...$) having the same (vector) quantum numbers as the photon. In consequence the bare electromagnetic vertex is multiplied by the following factor
\begin{equation}\label{3.1}
f_V= 1-\beta_V + \beta_V \left(\frac{m_V^2}{-k_V^2+m_V^2}\right)^n,
\end{equation} 
where $m_V$ and $k_V^2$ represent the mass and the 4-momentum transfer squared of the virtual vector meson, respectively. Also, $\beta_V$ represents the ``intensity'' of the vector meson coupling. In the last term of the previous equation we recognize straightforwardly the vector meson propagator. The expression of $f_V$ in (\ref{3.1}) is constructed in such way that, when the momentum transfer is vanishing, the charge normalization of the electromagnetic current is not affected, being $f_V(k_V^2=0)=1$.
\\ 
For the present example, unessential isospin operators are neglected.
\\ 
In standard VMD models, a monopole form $n=1$ is considered and $k_V^2$ is taken of the same magnitude as the virtual photon momentum transfer, that is
\begin{equation}\label{3.2}
k_V^2=q_\mu q^\mu = -\vec q^2_{(Breit)}.
\end{equation}
On the other hand, in the present work we assume that the state of the vector meson is determined by the dynamics of the hadronic constituent to which the vector meson is coupled, that is the \# 2. For this reason we take
\begin{equation}\label{3.3}
k_V^2=(p_{F,2}-p_{I,2})^2,
\end{equation}
that is the difference squared of the on shell momenta of the interacting constituent in the initial and final state; their explicit expression is obtained by means of equations (\ref{2.4}) and (\ref{2.5}). The same implementation of the VMD has been used to study the weak current of the nucleon in a relativistic quark model \cite{radici}.
\\ 
Multiplying $I^\mu$ by $f_V$ with the assumption of equation (\ref{3.3}) and performing the integration, we obtain the final result of the model.
\\ 
For the wave function we have taken a modified Hulten function given by the following expression
\begin{equation}\label{3.4}
\varphi(\vec p)= \frac{C}{\left(1+\frac{\vec p^2}{\alpha^2}\right)^a\left(1+\frac{\vec p^2}{\beta^2}\right)^b}.
\end{equation}
The values of the parameters $a$ and $b$ being $C$ the normalization constant, are chosen to reproduce the pion form factor and are given in the figure \ref{fig6}.
\\\\ 
In conclusion, this very simple scalar model has been used to study some good properties of the covariant impulse approximation introduced in \cite{impulse}. With the same model we tried to reproduce the pion form factor also introducing the effects of the VMD consistently with the other assumptions of the model.


\begin{thebibliography}{00}
   
\bibitem{impulse}
M. De Sanctis, M. A. Acero, D. A. Milanes, C. E. Sandoval, arXiv:nucl-th/0411088; to appear in Rev. Momento 28.

\bibitem{MDeS}
M. De Sanctis, M. M. Giannini, E. Santopinto and A. Vassallo, Rev. Mex. Fis. \textbf{50}S2, 96 (2004); also to appear in Nucl. Phys. A.

\bibitem{QFT}
C. Itzykson and J. B. Zuber, \textit{Quantum Field Theory}, Chapter 10, McGraw-Hill N.Y. (1988).
    
\bibitem{bebek}
C. J. Bebek \textit{et al}., Phys. Rev. \textbf{D17}, 1693 (1978).
    
\bibitem{volmer}
J. Volmer \textit{et al}., Phys. Rev. Lett \textbf{86}, 1713 (2001).

\bibitem{radici}
L. Ya. Glozman, M. Radici R. F. Wagenbrunn, S. Boffi, W. Klink, and W. Plessas, Phys. Lett \textbf{B516}, (2001)
   
\end{thebibliography}
\end{document}